\documentclass[%
a4paper,
prd,
10pt,
twocolumn,
superscriptaddress,
preprintnumbers,
nofootinbib,
nobibnotes,
amsmath,amssymb,
aps,
fleqn,
floatfix
]{revtex4-2}
\usepackage{amsfonts,graphics,color}
\usepackage{physics}
\usepackage{appendix}
\usepackage[english]{babel}
\usepackage{graphicx}
\usepackage{dcolumn}
\usepackage{bm}
\usepackage{xcolor}
\definecolor{xlinkcolor}{cmyk}{1,1,0,0}
\usepackage[bookmarks=true, pdfnewwindow=true, colorlinks=true, linkcolor=xlinkcolor, citecolor=xlinkcolor, filecolor=xlinkcolor, urlcolor=xlinkcolor, final=true]{hyperref}

\usepackage[normalem]{ulem}
\definecolor{myblue}{rgb}{0.05,0.1,0.5}

\begin{document}

\preprint{INR-TH-2024-023}

\title[Lorentz invariance violation and muons in showers]{Hypothetical Lorentz invariance violation and the muon content of extensive air showers}
\author{Nickolay\,S.\,Martynenko}
\affiliation{Lomonosov Moscow State University, 1-2 Leninskie Gory, Moscow 119991, Russia}
\affiliation{Institute for Nuclear
Research of the Russian Academy of Sciences, 60th October Anniversary Prospect 7a, Moscow 117312, Russia}
\author{Grigory\,I.\,Rubtsov}
\affiliation{Institute for Nuclear
Research of the Russian Academy of Sciences, 60th October Anniversary Prospect 7a, Moscow 117312, Russia}
\affiliation{Lomonosov Moscow State University, 1-2 Leninskie Gory, Moscow 119991, Russia}
\author{Petr\,S.\,Satunin}
\affiliation{Institute for Nuclear
Research of the Russian Academy of Sciences, 60th October Anniversary Prospect 7a, Moscow 117312, Russia}
\affiliation{Lomonosov Moscow State University, 1-2 Leninskie Gory, Moscow 119991, Russia}
\author{Andrey\,K.\,Sharofeev}
\email[Corresponding author: ]{sharofeev@inr.ac.ru}
\affiliation{Lomonosov Moscow State University, 1-2 Leninskie Gory, Moscow 119991, Russia}
\affiliation{Institute for Nuclear
Research of the Russian Academy of Sciences, 60th October Anniversary Prospect 7a, Moscow 117312, Russia}
\author{Sergey\,V.\,Troitsky}
\affiliation{Institute for Nuclear
Research of the Russian Academy of Sciences, 60th October Anniversary Prospect 7a, Moscow 117312, Russia}
\affiliation{Lomonosov Moscow State University, 1-2 Leninskie Gory, Moscow 119991, Russia}

\begin{abstract}
Extensive air showers (EAS), produced by cosmic rays in the atmosphere, serve as probes of particle interactions, providing access to energies and kinematical regimes beyond the reach of laboratory experiments. Measurements from multiple cosmic-ray detectors indicate a significant, yet unexplained, discrepancy between the observed muon content in EAS and that predicted by state-of-the-art interaction models, suggesting a need for refinements in our understanding of fundamental physics. Here we show that a tiny, experimentally allowed, violation of the Lorentz invariance (LIV) may result in the suppression of the number of electrons in EAS, leaving the muon number intact and explaining both the ``muon excess'' and its energy dependence. On the other hand, we use the lack of a much stronger discrepancy between EAS data and simulations to obtain strict constraints on the LIV scale. Future experimental tests of this LIV scenario are outlined. 
\end{abstract}

\maketitle


\section{Introduction}
\label{sec:Introduction}
One of the most intriguing problems in cosmic-ray physics is that of the ``muon excess'' in extensive air showers (EAS), or ``the muon puzzle''. According to data from a number of experiments, the muon content in EAS exceeds significantly the predictions of models of the development of the shower for the same energy of the primary particle. Solving the muon excess problem is a key to determining accurate models of the development of EAS. Understanding the EAS development is a necessary prerequisite for any reliable study of ultrahigh energy cosmic rays, and at the same time, EAS physics is a tool to study particle interactions in kinematical regimes hard to probe at accelerators. 

Apparently, the discrepancy between the numbers of muons and of electromagnetic particles in EAS was first noticed in the 1970s in the Sydney University Giant Air-shower Array (SUGAR) experiment \cite{mu1,mu2}, but the study of the EAS electromagnetic component was not a primary task for this facility, and the result was not widely discussed. In 2000, a joint study \cite{mu3} by High Resolution Fly's Eye (HiRes, which detected the electromagnetic component of the EAS using the fluorescence method) and Michigan Muon Array (MIA, which recorded the muon component of the same EAS by a ground-based array of detectors) collaborations reported a discrepancy in the number of muons in modeled and measured atmospheric showers in the range of primary particle energies from $10^{17}$ to $10^{18}$~eV. In 2008, a detailed analysis of the muon component of 33 individual events of the highest energy registered by the Yakutsk complex EAS facility \cite{mu4} indicated a 1.5-fold excess of muons in these showers compared to the SIBYLL model predictions. In 2010, the NEVOD-DECOR collaboration reported \cite{mu5} (more details in \cite{mu6}) an increase in the muon density in air showers compared to modeling. The problem of the muon excess began to attract the greatest attention after the publication of related results of the largest modern experiments, the Pierre Auger Observatory \cite{mu7,mu:newAuger} and Telescope Array~\cite{mu9}. Recent results of KM3NeT \cite{mu:KM3NeT}, together with new analyses of data from the discontinued SUGAR \cite{mu10,mu11} and Akeno Giant Air Shower Array, AGASA, \cite{mu12} experiments, confirm the presence of muon excess. At the same time, in a number of experiments, the excess of muons in the observed EAS compared to the model one was not detected. They include the Moscow State University EAS array, EAS-MSU \cite{mu13}; Karlsruhe Shower Core and Array Detector (KASCADE) Grande \cite{mu14}; IceTop \cite{mu15}; Haverah Park in the recent reanalysis \cite{mu16}; and, at not the highest energies, the Yakutsk experiment \cite{mu17}. A discussion of the experimental status of the muon excess problem can be found in the review \cite{mu18}.

Answering the key question of the nature of the muon excess in EAS would require a joint analysis of the data of different experiments. To reach these goals, the Working Group on Hadronic Interactions and Shower Physics (WHISP) was established. The results of this meta-analysis so far indicate that there is no direct contradiction between the different experiments, since each of them operates in its own range of primary particle energies, zenith angles of EAS arrival, energies of detected muons, at its own atmospheric depth, etc. The combined data from nine independent experiments were found to indicate the presence of a muon excess increasing with the primary particle energy \cite{mu20,mu21,ArteagaVelazquez:2023fda}. The muon excess is often parametrized by the variable
\begin{equation}
    z \equiv \frac{\ln \expval{N_\mu^{\text{obs}}} - \ln \expval{N_{\mu, p}^{\text{MC}}}}{\ln \expval{N_{\mu, Fe}^{\text{MC}}} - \ln \expval{N_{\mu, p}^{\text{MC}}}},
    \label{eq:z-scale}
\end{equation}
where $\expval{N_\mu^{\text{obs}}}$ is the mean value of the measured muon density, $\expval{N_{\mu, p}^{\text{MC}}}$ and $\expval{N_{\mu, Fe}^{\text{MC}}}$ are the values for the average muon density for proton and iron cosmic-ray nuclei, respectively, as predicted by Monte-Carlo (MC) simulations.

Despite numerous attempts to explore possible modifications of hadronic interactions in the air shower, a satisfactory theoretical description of both the origin of the muon excess and of its dependence on experimental conditions, consistent with collider results, has not yet been constructed  \cite{mu7,mu18,Ostapchenko:2024xbe,mu:new-kabak}. Here we follow a completely different approach, which leaves the hadronic interactions intact but modifies the electromagnetic interactions in an experimentally allowed way. Namely, we invoke new-physics models with Lorentz-invariance violation (LIV). 

The idea of LIV comes from several approaches to quantum gravity, see \cite{Addazi:2021xuf} for a review, and can be tested experimentally at high energies. LIV models predict rich phenomenology manifesting itself in astrophysical processes: appearance of new  particle decay channels forbidden in Lorentz-invariant (LI) case and shifts of thresholds for particle-interaction processes. One of the LIV predictions of our current interest is the suppression of photon pair production in the Coulomb field, that is of the Bethe--Heitler process \cite{Bethe:1934za}. It leads to the suppression of formation of photon-initiated EAS \cite{Vankov:2002gt, Rubtsov:2012kb} for a specific type of LIV in which photon has negative energy-dependent effective mass squared (subluminal scenario); the sensitivity of the corresponding method to the LIV mass-scale parameter grows with energy. This phenomenon has been previously investigated, and corresponding constraints have been established \cite{Rubtsov:2016bea, Satunin:2019gsl, Satunin:2021vfx}, based on the detection of EAS initiated by primary photons with energies up to $\sim 1$ PeV. Besides, an EAS initiated by a primary hadron with energy $\sim 10^{19}$~eV includes secondary photon-initiated subshowers \footnote{The idea to test LIV in electromagnetic sector with hadron EAS was proposed in \cite{Klinkhamer:2017puj} in the context of a different scenario.}. Energies of these secondary photons are $\sim  10^{17}$ eV, two orders of magnitude higher than the energy of primary photons ever detected. Therefore, the corresponding values of LIV parameters are still not experimentally tested. In the case of LIV of this type, photon subshowers would become deeper and would produce fewer electrons. Hence the energy of the primary particle, reconstructed within a LI model, would be smaller than its true energy. 
The number of muons in both LI and LIV cases remains the same for a fixed \textit{true} energy of primary hadron. However, for a fixed \textit{reconstructed} primary energy, the number of muons in the LIV case is larger. 
This is exactly what appears as the muon puzzle. Below, we test this idea more accurately by performing MC shower simulations of hadron-induced EAS with a modified Bethe--Heitler cross section for secondary photons, assuming LIV models for photons with various values of the LIV mass scale. 

Traditional bounds on LIV come from searches of energy-dependent time lags for electromagnetic signals from distant sources, see e.g.\  \cite{HESS:2011aa, Vasileiou:2013vra}, while stronger constraints come from the lack of cross-section modifications. The superluminal scenario (positive energy-dependent effective photon mass squared) implies some channels of photon decay, $\gamma \to e^+e^-$ and $\gamma\to 3\gamma$.  The corresponding bounds on the superluminal LIV are in general stronger \cite{Addazi:2021xuf} than on the subluminal one, on which we concentrate here.

An important process which is sensitive to LIV of the subluminal type is the pair production, $\gamma\gamma_b \to e^+e^-$, by an energetic photon $\gamma$ on a soft background photon $\gamma_b$. In the LI scenario, it results in the attenuation of extragalactic photons beyond TeV energies due to the pair production on cosmic infrared and microwave background radiation. Subluminal LIV shifts the threshold of the process, making extragalactic photon flux almost unattenuated, see \cite{Kifune:1999ex, Stecker:2001vb}  and \cite{Martinez-Huerta:2020cut} for review. The current attenuation constraints \cite{Lang:2018yog} are an order of magnitude weaker than Bethe--Heitler ones \cite{Satunin:2021vfx}, see \cite{Addazi:2021xuf} for review.

Different types of LIV were considered in previous works on EAS simulations. Ref.~\cite{Antonov:2001xh} performed MC simulations of hadronic EAS assuming LIV resulting in non-decaying $\pi_0$s which lead to a decrease of the depth of the maximal shower development. The authors of \cite{Diaz:2016dpk,Klinkhamer:2017puj,Duenkel:2021gkq} performed MC simulations of hadronic EAS in the case of superluminal LIV, when the $\gamma\to e^+e^-$ process is allowed, and put the corresponding constraints. Additionally, the modification of pion decay \cite{Klinkhamer:2016evw} and the vacuum \u{C}erenkov process \cite{Duenkel:2023nlk} in EAS formation have been taken into account. 
However, the simulations of EAS with modified Bethe-Heitler process  
have not been performed until now. 
  
The rest of the paper is organized as follows. Section~\ref{sec:Theory} introduces the effective field theory (EFT) approach to the formulation of LIV models, which we use throughout the paper. In Sec.~\ref{sec:Muon puzzle}, we demonstrate how to solve the muon puzzle with LIV EFT. Sec.~\ref{sec:MC} presents simulations of EASs based on our LIV model. In Sec.~\ref{sec:MaxLik}, we analyze the results and set data-driven constraints on the LIV parameter $M_{\text{LIV}}$ in the context of our study. We outline our conclusions in Sec.~\ref{sec:concl}.

\section{Lorentz Invariance violation}
\label{sec:Theory}
LIV is usually parametrized either by modified dispersion relations or by Effective Field Theory (EFT).
The latter approach is more self-consistent from the point of view of quantum field theory, whereas the former may represent a superficial phenomenological description that does not reproduce all the symmetries of the original theory. Modified dispersion relation for photon reads
\begin{equation}
    E_\gamma^2 = k_\gamma^2+\sum_{n=1,2,..}s_n\frac{k_\gamma^{2+n}}{M_{\text{LIV},(n)}^n},
\end{equation} 
where $E_\gamma$ and $k_\gamma$ denote photon energy and momentum, $M_{\text{LIV},(n)}$ is the LIV mass scale, $s_n=\pm 1$ is a sign. The most studied cases are $n=1$ and $n=2$. The EFT corresponding to $n=1$, the Myers-Pospelov model \cite{Myers:2003fd}, implies the opposite signs $s_1$ for different photon polarizations, which lead to the birefringence phenomenon and, in turn, to very strong constraints \cite{Gotz:2013dwa} on the scale, many orders of magnitude larger than the shower formation ones \cite{Satunin:2023yvj}. 
In this article, we consider the $n=2$ case, which is also motivated by Horava-type models of quantum gravity \cite{Eichhorn:2019ybe}. 

\subsection{An EFT approach to $n=2$ LIV Quantum Electrodynamics}
\label{subsec:EFT}
To study $n=2$ LIV Quantum Electrodynamics (QED), we follow the EFT approach \cite{Mattingly:2008pw,Rubtsov:2012kb}. We consider QED with additional LIV EFT operators of dimensions up to $6$, see additional requirements in \cite{Rubtsov:2012kb},
\begin{equation}
    \mathcal{L} = \mathcal{L}_{\text{QED}} + \mathcal{L}_e + \mathcal{L}_\gamma,
    \label{eq:lagrangian}
\end{equation}
where
$ \mathcal{L}_{\text{QED}}$ is the standard QED Lagrangian.
The LIV operators for photons and electrons are
\begin{align}
  &  \mathcal{L}_\gamma = \frac{s_2}{4 M_{\text{LIV}}^2} F_{k j} \partial_i^2 F^{k j}, 
    \label{eq:photon's part} \\
  &    \mathcal{L}_e = i \kappa \bar{\psi} \gamma^i D_i \psi + \frac{i s_2^e}{M_{\text{LIV, e}}^2} D_j \bar{\psi} \gamma^i D_i D_j \psi,
    \label{eq:fermion's part} 
\end{align}
respectively, where the covariant derivative is defined as $D_\mu \psi = \pqty{\partial_\mu + i e A_\mu} \psi$, $A_\mu$ is the electromagnetic field strength, $F_{\mu\nu}$ is its stress tensor, and $\psi$ is the electron field. Here $M_{\text{LIV}}$ and $M_{\text{LIV, e}}$ are mass scales for LIV in the photon and electron sectors (which, in general, are different), $s_2$ and $s_2^{(e)}$ are $\pm 1$ and $\kappa$ is a dimensionless parameter. 

We consider a subluminal LIV scenario in which the photon dispersion relation is modified as follows,
\begin{equation}
    E_\gamma^2 = k_\gamma^2 - \frac{k_\gamma^4}{M_{\text{LIV}}^2}.
    \label{eq:Dispersion relation}
\end{equation}

In this paper, we focus exclusively on the effects arising from $\mathcal{L}_\gamma$. The reason for neglecting $\mathcal{L}_e$ is the following: 
the constraints on $M_{\text{LIV, e}}$ came from the absence of anomaly synchrotron and vacuum Cerenkov radiation of soft photons by electrons in Crab Nebula, are two-sided and exclude the mass scale $M_{\text{LIV, e}} = 2 \, \times \,10^{16}$~GeV at $95\,\%$ CL \cite{Liberati:2012jf}, while the current constraints on $M_{\text{LIV}}$  in the subluminal scenario \cite{Satunin:2021vfx} is  
\begin{equation}
\label{eq:constr}
    M_{\text{LIV}} > 1.7\times 10^{13} \; \text{GeV}. 
\end{equation}
Thus, analyzing LIV for mass scales $10^{(13\dots16)}$ GeV we can neglect LIV for electrons. In this paper, we neglect LIV for electrons even for higher mass scales since we propose a particular LIV solution of muon puzzle which can be generalized to account for electron LIV parameters.




\subsection{The Bethe--Heitler process in the context of EAS}
\label{subsec:BH process}
The Bethe--Heitler process is the most probable channel of the first interaction of a photon in the atmosphere. The LI result for the cross section $\sigma^{\text{LI}}_{\text{BH}}$ of this process was first calculated in~\cite{Bethe:1934za},
\begin{equation}
    \sigma^{\text{LI}}_{\text{BH}} = \frac{28 Z^2 \alpha^3}{9 m_e^2} \bqty{\log \frac{183}{Z^{1/3}} - \frac{1}{42}},
\end{equation}
where $\alpha$ is the fine-structure constant, $m_e$ is the electron mass, and $Z$ is the nucleus' charge. This result includes the screening effect in an atom. In the LIV case, the cross section is modified as shown in~\cite{Rubtsov:2012kb, Rubtsov:2016bea}, and the result in the high--energy limit is given by
\begin{equation}
    \label{eq:LIV-suppression}    \frac{\sigma^{\text{LIV}}_{\text{BH}}}{\sigma^{\text{LI}}_{\text{BH}}} \simeq \frac{12 m_e^2 M^2_{\text{LIV}}}{7 E_\gamma^4} \times \log{\frac{E_\gamma^4}{2 m_e^2 M^2_{\text{LIV}}}}
\end{equation}
in the limit $E_\gamma^4/\pqty{2 m_e^2 M^2_{\text{LIV}}} \gg 1$. The most important implication is that the cross section decreases as $E_\gamma^{-4} \log E_\gamma^4$, leading to fewer electrons born in high-energy interactions compared to the LI case. In the context of EAS, this results in an increased average propagation length of high-energy photons produced during EAS evolution. Consequently, fewer electron–positron pairs are produced compared to the LI scenario.

\subsection{Other processes. Vacuum Cerenkov radiation}

Other processes become allowed or get modified in the case of subluminal LIV. If a photon is subluminal, even a LI electron can emit a photon in a so-called Vacuum Cerenkov process \cite{Jacobson:2002hd,Konopka:2002tt,Jacobson:2005bg}. In the case of dispersion relation (\ref{eq:Dispersion relation}), the process in which an electron almost stops  emitting a hard photon, has the threshold
\begin{equation}
    E_e > \left(2m_e M_{\text{LIV}}^2\right)^{1/3}.
    \label{eq:10}
\end{equation}
The limit  (\ref{eq:constr}) and Eq.~(\ref{eq:10}) imply that, for electrons with energies $E_e < 10^{17}$ eV, the vacuum Cerenkov radiation does not occur. The contribution of higher-energy electrons to the development of EAS initiated by primaries of energies $E\lesssim 10^{19}$ eV is negligible, so we do not consider the vacuum Cerenkov process in our analysis.

\section{Solution to the muon puzzle}
\label{sec:Muon puzzle}
Let us demonstrate how the EFT with LIV could solve the muon puzzle. Consider an EAS induced by an ultra-high energy cosmic-ray particle (e.g. proton with \(E \sim 10^{19}\)~eV). The first interaction occurs almost immediately after the particle enters the atmosphere, leading to the production of charged and neutral pions (\(\pi^\pm\), \(\pi^0\)). The charged pions subsequently interact with the atmosphere,
while the neutral pions decay into two photons~\cite{2005Heitler_Matthews}. 

We are particularly interested in these high-energy photon pairs, which induce electromagnetic sub-showers, contributing significantly to the ground-level lepton content of the EAS. 
Within the LIV model, the cross section of the Bethe--Heitler process, \(\sigma_{\rm{BH}}\), is suppressed  at high energies (\(\sigma_{\text{BH}}^{\text{LIV}} \ll \sigma_{\text{BH}}^{\text{LI}}\)), leading to an increased photon mean free path \(\lambda\) in the atmosphere, \(\lambda^{\text{LIV}} \gg \lambda^{\text{LI}}\), and, in fact, eliminating these electromagnetic sub-showers.
As a result, the average number \(\expval{N_e}\) of electrons reaching the detector is lower in the case of LIV, \(\expval{N_{e,\text{LIV}}} < \expval{N_{e,\text{LI}}}\).

The key moment of the proposed solution to the muon puzzle is that the primary energy reconstruction is often based on measuring $N_e$ in one way or another. Indeed, modern hybrid experiments, Pierre Auger Observatory \cite{AugerEnergy} and Telescope Array \cite{TAenergy}, calibrate their energy measurements with the fluorescent detector (FD), and the air fluorescence is determined by the shower electrons. Since the electrons carry most of the energy in the shower, the method is often called calorimetric. However, this approach assumes some missing-energy correction due to the EAS muons which is estimated either from MC simulations (in Telescope Array), or from indirect measurements of the muon excess \cite{AugerMissingE} (in Pierre Auger Observatory). In meta-analyses \cite{ArteagaVelazquez:2023fda}, energy scales of various experiments were adjusted to study the energy dependence of the muon excess. Going into details of the energy reconstruction of each particular experiment is far beyond the scope of this paper. To keep our idea more transparent, and without loss of generality, we assume that the reconstructed energy \(E_{\text{reco}}\) and the reconstructed mass number \(A_{\text{reco}}\) of a EAS primary particle is related to \(\expval{N_e}\) by a simple power-law scaling, \(\expval{N_e} \propto A^{-\alpha_e}_{\text{reco}}E^{\beta_e}_{\text{reco}}\):
\begin{equation}
    \label{eq:Ereco-Ne-relation}
    \ln [E_{\text{reco}} /\text{GeV}] = \varepsilon_{e} + (\alpha_e\beta^{-1}_{e}) \ln A_{\text{reco}} + \beta^{-1}_{e} \ln \expval{N_e}.
\end{equation}

Parameters \(\varepsilon_e\), \(\alpha_e\) and \(\beta_e\) of Eq.~\eqref{eq:Ereco-Ne-relation} are determined with the help of either MC simulations, using that the analogous scaling,
\begin{equation}
    \label{eq:Etrue-Ne-relation}
    \ln [E / \text{GeV}] = \varepsilon_{e} + (\alpha_e\beta^{-1}_{e}) \ln A + \beta^{-1}_{e} \ln \expval{N_e},
\end{equation}
is applicable, or data-driven methods, assuming the LI theory. \(A_{\text{reco}}\), in fact, encodes any additional variables other than \(\expval{N_e}\), which are used for energy reconstruction in the actual experimental data analysis.

Since, for the LIV model we study here, 
\(\expval{N_{e,\text{LIV}}} < \expval{N_{e, \text{LI}}}\), this may lead to a bias in the energy reconstruction, \(E_{\text{reco}} < E\). 
On the other hand, our LIV model has a negligible impact on the number of muons, since the latter are predominantly produced by charged pion decays in the energy range of \(E_{\pi^{\pm}} \lesssim 10^{11}\)~eV, see e.g.~\cite{2005Heitler_Matthews} and references therein. 
We justify this explicitly with our simulations, see Section~\ref{sec:MC}.

Analogously, we adopt a simple power-law scaling \(N_{\mu} \propto A^{\alpha_\mu}E^{\beta_\mu}\)
\begin{equation}
\label{eq:Nmu-Etrue-relation}
\ln\expval{N_\mu} = -n_{\mu} +\alpha_\mu \ln A+ \beta_{\mu} \ln[E /\text{GeV}].
\end{equation}
Here, the opposite sign of \(\alpha_\mu\) compared to \(\alpha_e\) is due to the natural expectation that the larger mass number \(A\), the greater the fraction of \(E\) is released in a hadronic cascade, where muons are eventually produced, and hence, the smaller the fraction of \(E\) converted to an electromagnetic cascade~\cite{PhysRevD.46.5013}.

In our analysis, we identify \(\expval{N_{e, \text{LIV}}}\) with the observed \(\expval{N_{e}}\) and estimate the reconstructed energy according to Eq.~\eqref{eq:Ereco-Ne-relation}. For the LIV EASs, we identify the expected values of \(\expval{N^{\text{MC}}_{\mu, p}}\) and \(\expval{N^{\text{MC}}_{\mu, Fe}}\) with those predicted by Eq.~\eqref{eq:Nmu-Etrue-relation} for the reconstructed energy \(E_{\text{reco}}\) instead of the true primary energy \(E\).

Therefore, the actual number of muons produced in the LIV EAS is underestimated. This induces larger-than-expected values of \(z\), see Eq.~\eqref{eq:z-scale}, but due to “lack of electrons” and subsequent biased energy reconstruction rather than due to “excess of muons”,
\begin{equation}
    \label{eq:z-scale-model}
    z = \frac{\ln A}{\ln 56} + \frac{\beta_\mu}{\alpha_\mu\ln 56}\ln\left[\frac{E}{E_{\text{reco}}}\right].
\end{equation}
For conciseness, we hereafter use the notation 
\begin{equation}
    r_e \equiv \ln\left[\frac{\expval{N_{e,\text{LI}}}}{\expval{N_{e,\text{LIV}}}}\right],\quad r_\mu \equiv \ln\left[\frac{\expval{N_{\mu,\text{LI}}}}{\expval{N_{\mu,\text{LIV}}}}\right].
\end{equation}

The energies of the first-generation photons are expected to scale nearly linearly with the energy per nucleon, \(E/A\), of the primary particle. Taking into account Eq.~\eqref{eq:LIV-suppression}, we conclude that  the universal parameter describing the magnitude of the LIV effects in our study is
\begin{equation}
    \xi \equiv (m_e M_{\text{LIV}})^{-1/2}A^{-1}E.
\end{equation}
For further analysis, we adopt a simple parametrization which reflects well the trend of \(r_e\) and \(r_\mu\) seen in the MC simulations:
\begin{equation}
    \label{eq:r-xi}
    r_e(\xi) = r_{e,0} \ln\left[1 + \left(\frac{\xi}{\xi_0}\right)^{\varrho}\right],\quad r_\mu(\xi) = 0,
\end{equation}
where \(r_{e, 0}\), \(\xi_0\), and \(\varrho\) are free positive parameters. We note that such a parametrization guarantees a correct LI limit: \(\lim\limits_{\xi \rightarrow 0}r_e(\xi) = 0\). 

\section{Monte-Carlo simulations}
\label{sec:MC}
In our MC simulations, we utilize CORSIKA version 7.7550~\cite{1998CORSIKA_Physics}, incorporating the EPOS 1.99 (UrQMD 1.3.1) model for high-energy (low-energy) hadronic interactions and EGS4 for electromagnetic interactions. We always model vertical (zenith angle \(\theta = 0\)) EASs with a fixed primary particle energy and assume detection at the sea level. To reduce the computational time, we enable the thinning option~\cite{corsika_user_guide} with the parameter \(\epsilon_{\text{thin}} = 10^{-2}\) and with the weight limitation according to~\cite{Kobal:2001jx}. We use the maximum weight value of \(w_{\max}= 10^8\). The weight ratio for the electromagnetic and hadronic particles is set to 10. Since throughout this study we are primarily interested in number-of-particle ratios, particular simulation parameters are not expected to significantly affect our results and qualitative conclusions. 

We model only proton--induced (\(A=1\)) and iron nucleus--induced (\(A=56\)) EASs. It will be further shown that this is sufficient for the purposes of our analysis, which is insensitive to the primary cosmic-ray composition.

In all simulations, we set the energy threshold of \(E_e \geq 1\)~MeV for electrons and \(E_\mu \geq 1\)~GeV for muons. 

At the first step, we simulate air showers without any LIV to reproduce the \(\expval{N_e}\)--based energy reconstruction and the energy--based \(\expval{N_\mu}\) estimation. We fit \(\varepsilon_e\), \(\alpha_e\), \(\beta_e\), \(n_\mu\), \(\alpha_\mu\), and \(\beta_\mu\) parameters, see Eqs.~\eqref{eq:Ereco-Ne-relation}--\eqref{eq:Nmu-Etrue-relation}, to MC simulation results in the primary particle energy range of \(10^{16}\)~eV \(\leq E \leq\) \(5.0\times 10^{19}\)~eV. We average the number of particles at the ground level over 100 EAS simulations for each pair \((A, E)\), and fit a weighted linear regression, inducing weights from the \(N_e\) standard deviations. The results are presented in Table~\ref{tab:LinearRegressionResults}.

\begin{table}[htb]
    \centering
    \begin{tabular}{c|c|c|c|c|c|c}
        \textbf{parameter} & \(\varepsilon_e\) & \(\alpha_e\beta_e^{-1}\) & \(\beta_e^{-1}\) & \(n_\mu\) & \(\alpha_\mu\) & \(\beta_\mu\) \\
        \hline\hline
        best-fitting value & \(3.832\) & \(0.089\) & \(0.890\) & \(3.621\) &  \(0.076\) & \(0.921\)\\
        \hline
        standard error & \(1.169\) & \(0.046\) & \(0.053\) & \(0.173\) &  \(0.012\) & \(0.008\)
    \end{tabular}
    \caption{The best-fitting parameters of Eqs.~\eqref{eq:Ereco-Ne-relation}--\eqref{eq:Nmu-Etrue-relation} and their \(68\%\) C.L. uncertainties.}
    \label{tab:LinearRegressionResults}
\end{table}

The root mean squared errors of our linear models are:
\begin{equation}
    \label{eq:RMSE-for-linreg}
    \begin{aligned}
    \expval{\delta \ln [E_{\text{reco}}/\text{GeV}]^2}^{1/2} &\simeq 0.085,\\
    \expval{\delta \ln \expval{N_{\mu}}^2}^{1/2} &\simeq 0.021.
    \end{aligned}
\end{equation}
Then, we modify the EGS4 Bethe--Heitler cross-section (see Appendices \ref{sec:appendix:EM}, \ref{sec:appendix:BH}) in order to reproduce the suppression described in Subsection~\ref{subsec:EFT}. We scan over the LIV mass scale \(M_{\text{LIV}} \in \{10^{13}, 10^{14}, \dots, 10^{18}\}\)~GeV, and vary the thrown energy 
for each \(M_{\text{LIV}}\). 

Using this data set, which we call training, we fit \(r_e(\xi)\) to the model~\eqref{eq:r-xi}, minimizing the weighted mean squared error. Table~\ref{tab:ratio-xi-scaling-results} presents the best-fitting parameters together with their uncertainties. We also test whether \(r_\mu(\xi)=0\) holds. 

\begin{figure}[htb]
    \centering
    \includegraphics[width=0.95\columnwidth]{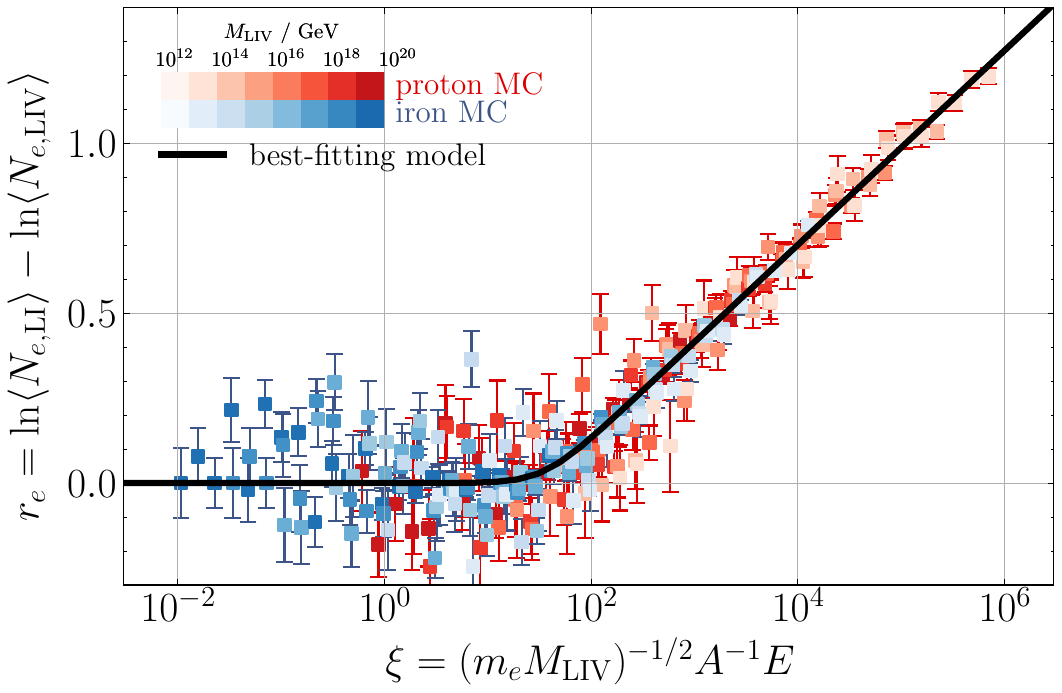}\\
    \vspace{\baselineskip}
    \includegraphics[width=0.95\columnwidth]{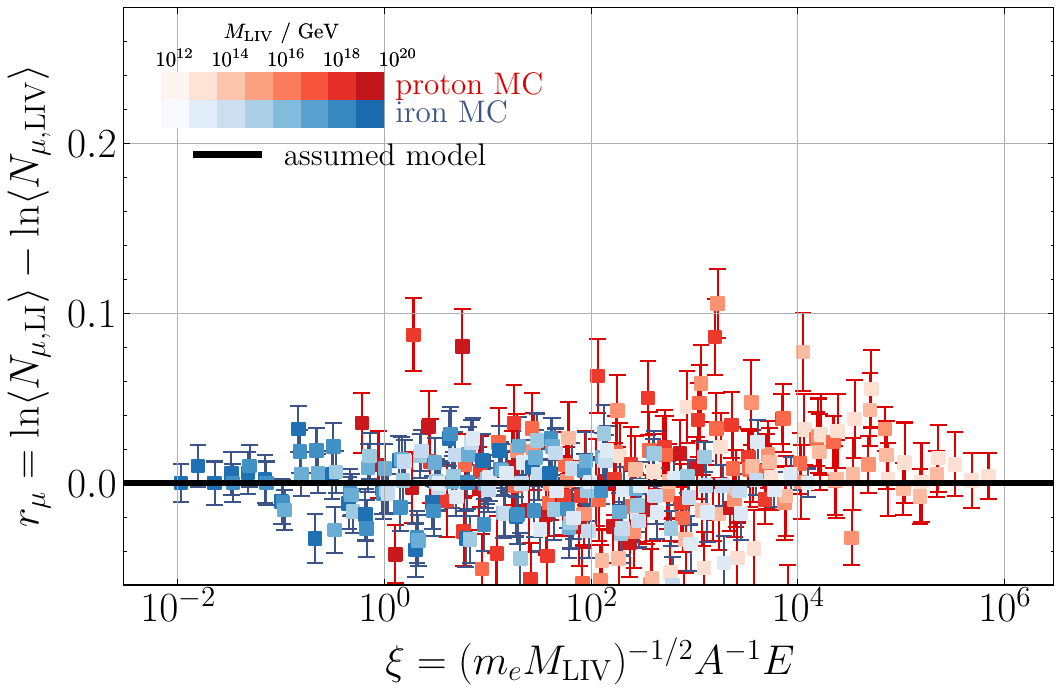}
    \caption{LI--to--LIV  ratios for the ground-level average number of particles, training data set (top: electrons, bottom: muons). The colored points and the error bars correspond to MC simulations (red: proton-induced, blue: iron-induced EASs) average and \(68\%\)~C.L. uncertainty, respectively. The solid black lines represent the model defined by Eq.~\eqref{eq:r-xi}. See text for a detailed discussion.}
    \label{fig:ratios-train}
\end{figure}

\begin{table}[htb]
    \centering
    \begin{tabular}{c|c|c|c}
        \textbf{parameter} & \(r_{e, 0}\) & \(\xi_0\) & \(\varrho\) \\
        \hline\hline
        best-fitting value & 0.052 & 35.290 & 2.407 \\
        \hline
        standard error & 0.013 & 2.156 & 0.568
    \end{tabular}
    \caption{The best-fitting parameters of Eq.~\eqref{eq:r-xi} and their \(68\%\) C.L. uncertainties.}
    \label{tab:ratio-xi-scaling-results}
\end{table}
For both \(r_e(\xi)\) and \(r_\mu(\xi)\) we obtain that the root of weighted mean squared error on a training dataset is approximately \(1.5\), which is acceptable within the scope of this work. Figure~\ref{fig:ratios-train} compares the model predictions with the training data.

\begin{figure}[htb]
    \centering
    \includegraphics[width=0.95\columnwidth]{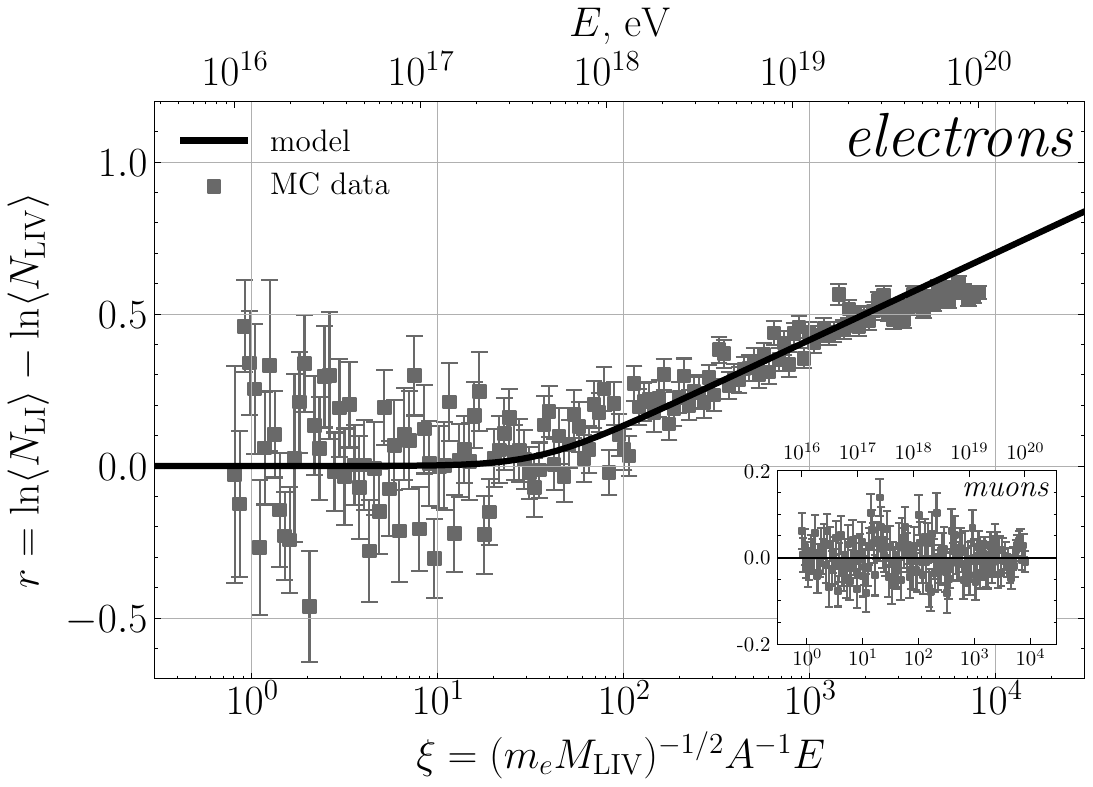}\\
    \vspace{\baselineskip}
    \includegraphics[width=0.95\columnwidth]{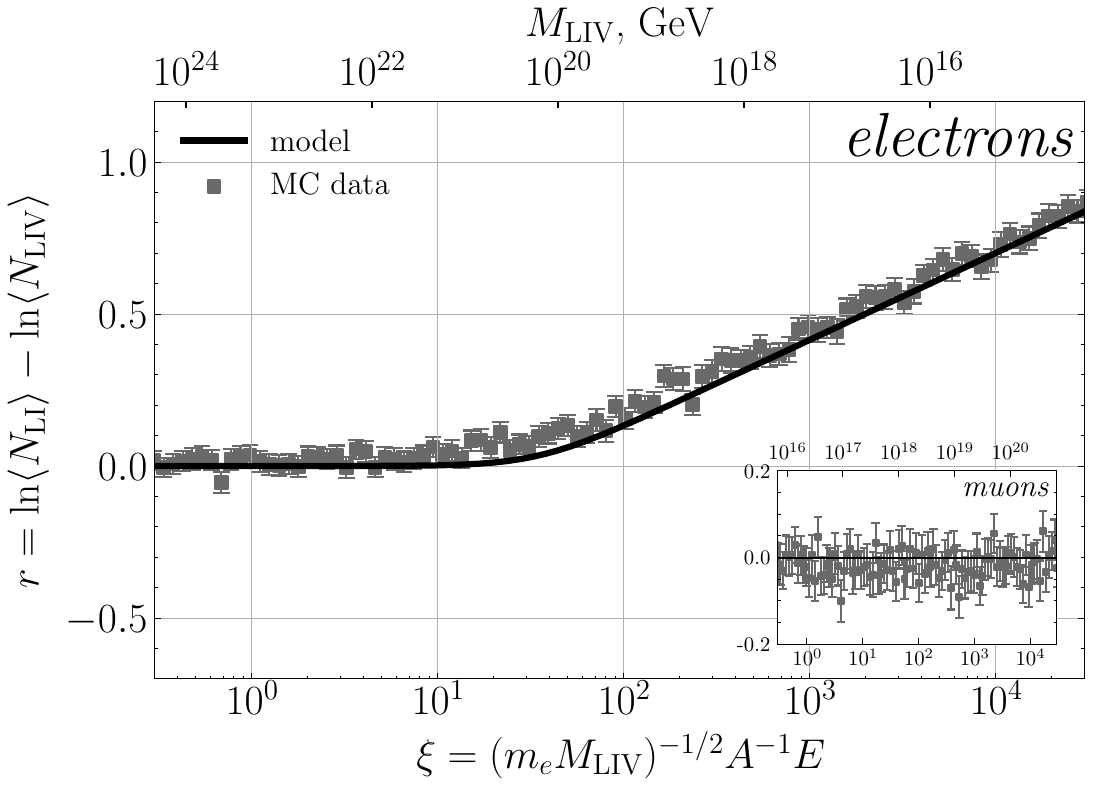}
    \caption{LI--to--LIV  ratios for the ground-level average number of particles, test data set (main figures: electrons, inset figures: muons). The gray points and the error bars correspond to MC simulations (proton-induced EASs) average and \(68\%\)~C.L. uncertainty, respectively. The solid black lines represent the model defined by Eq.~\eqref{eq:r-xi}. Top: fixed \(M_{\text{LIV}} = 3\times 10^{17}\)~GeV, bottom: fixed \(E=10^{19}\)~eV. See text for a detailed discussion.}
    \label{fig:ratios-test}
\end{figure}

To test the fitted model \(r_e(\xi)\) and additionally justify the hypothesis \(r_{\mu}(\xi) = 0\), we perform supplementary MC simulations of proton-induced EASs: (a) with fixed \(M_{\text{LIV}}=3\times 10^{17}\)~GeV and varying primary energy \(10^{16}\)~eV \(\leq E \leq 5\times 10^{19}\)~eV, and (b) with fixed \(E=10^{19}\)~eV and varying \(10^{14}\)~GeV \(\leq M_{\text{LIV}}\leq 10^{25}\)~GeV. Figure~\ref{fig:ratios-test} compares the best-fitting model predictions with the test data set described above.

One can see that the MC simulations explicitly justify the use of Eqs.~\eqref{eq:Ereco-Ne-relation}--\eqref{eq:z-scale-model}, and \eqref{eq:r-xi}. The actual particle ratios trend is well captured by our parametric model.

\section{Maximum Likelihood Analysis}
\label{sec:MaxLik}
We introduce \(\xi_{\text{reco}}\equiv (m_e M_{\text{LIV}})^{-1/2} A^{-1} E_{\text{reco}}\) and, utilizing Eqs.~\eqref{eq:Ereco-Ne-relation} and~\eqref{eq:Etrue-Ne-relation}, obtain:
\begin{equation}
    \label{eq:implicit-func}
    \ln \xi - \beta_e^{-1} r_e(\xi) = (\alpha_e \beta_e^{-1})\ln \left[\frac{A}{A_{\text{reco}}}\right] + \ln \xi_{\text{reco}}.
\end{equation}
One can solve this equation for \(\ln \xi\), for instance, employing the Newton--Raphson method with an initial guess \(\ln \xi=\ln \xi_{\text{reco}}\). This induces an implicit function \(E(E_{\text{reco}}, M_{\text{LIV}}, A, A_{\text{reco}})=(m_e M_{\text{LIV}})^{1/2} A \,\xi^\star\), where \(\xi^\star\) is the root of  Eq.~\eqref{eq:implicit-func}. Combining this with Eq.~\eqref{eq:z-scale-model}, we conclude that
\begin{equation}
    z(E_{\text{reco}} |  M_{\text{LIV}}, A, A_{\text{reco}}) = \frac{\ln A}{\ln 56} + \frac{\beta_\mu\ln\left[\xi^\star/\xi_{\text{reco}}\right]}{\alpha_\mu\ln 56}.
    \label{eq:z_Ereco_relation}
\end{equation}
To constrain \(M_{\text{LIV}}\), we fit the model \eqref{eq:z_Ereco_relation} to the Pierre Auger Observatory (Auger) data \(\{(E_{\text{reco},i}, z_i, \sigma_{z, i})\}_{i=1}^N\) presented in~\cite{ArteagaVelazquez:2023fda}, which is a combined analysis of fluorescence detector and surface detector data (referred to as FD+SD), as well as a combined analysis of underground muon detector and surface detector data (referred to as UMD+SD)~\cite{AugerZ}. The error of primary composition reconstruction has already been taken into account in the uncertainty \(\sigma_z\). Hence, in our model, we assume a perfect primary composition reconstruction \(A_{\text{reco}}=A\). Since we are interested in a composition-independent constraint, we define:
\begin{equation}
    \label{eq:chi-2}
    \chi^2 \equiv \sum_{i=1}^N \min\limits_{A\in[1, 56]}\left[\frac{z_i - z(E_{\text{reco}, i} | M_{\text{LIV}}, A, A)}{\sigma_{z, i}}\right]^2
\end{equation}

Next, we use the one-sided \(\chi^2\) test with \(N-1=12\) degrees of freedom to set a 95\% C.L. constraint on \(M_{\text{LIV}}\).

\begin{figure}[htb]
    \centering
    \includegraphics[width=0.95\columnwidth]{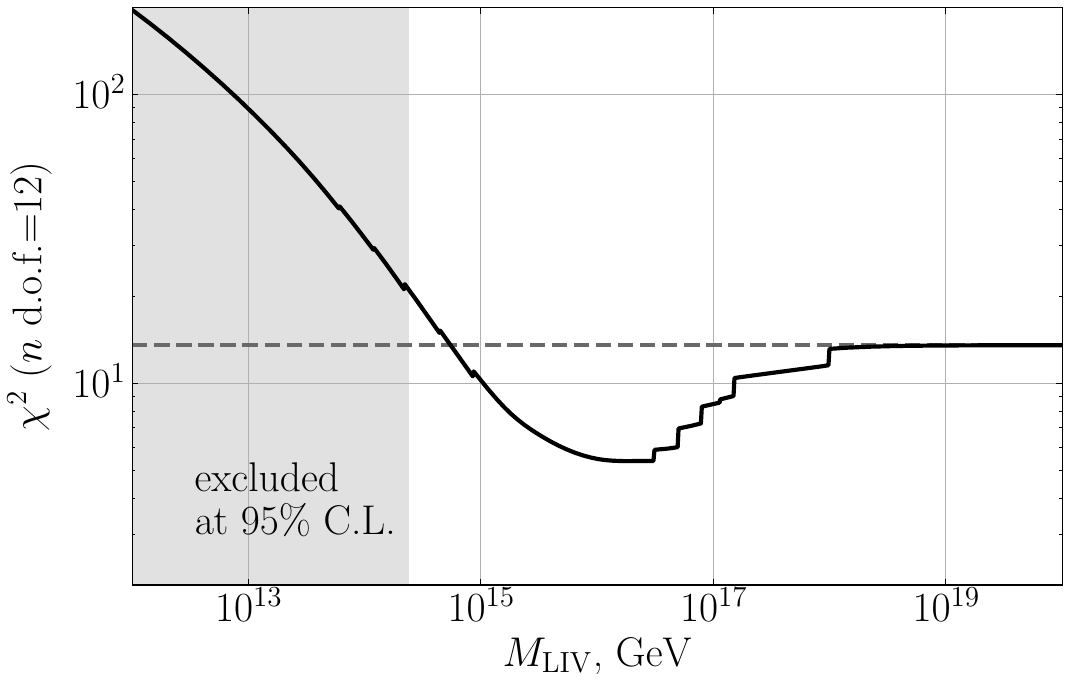}
    \caption{The solid black line corresponds to \(\chi^2\) at various \(M_{\text{LIV}}\), see~Eq.~\eqref{eq:chi-2}. The dashed gray line shows the \(\chi^2\) value for the LI scenario. The shaded region represents the parameter range excluded at \(95\%\) C.L.}
    \label{fig:chi2-analysis}
\end{figure}

The result is shown in Figure~\ref{fig:chi2-analysis}. One can see that the Auger data excludes \(M_{\text{LIV}} \leq 2.4 \times 10^{14}\)~GeV at 95\% C.L. We obtain that the most favorable LIV model in terms of the maximum-likelihood (i.e. minimum-\(\chi^2\)) analysis has \(M_{\text{LIV}}=1.9\times 10^{16}\)~GeV. In addition, we find that \(M_{\text{LIV}} \geq 6.3\times 10^{19}\)~GeV would make the LIV effect completely unrevealed in observations in terms of \(z\)-scale.

\begin{figure}[htb]
    \centering
    \includegraphics[width=0.95\columnwidth]{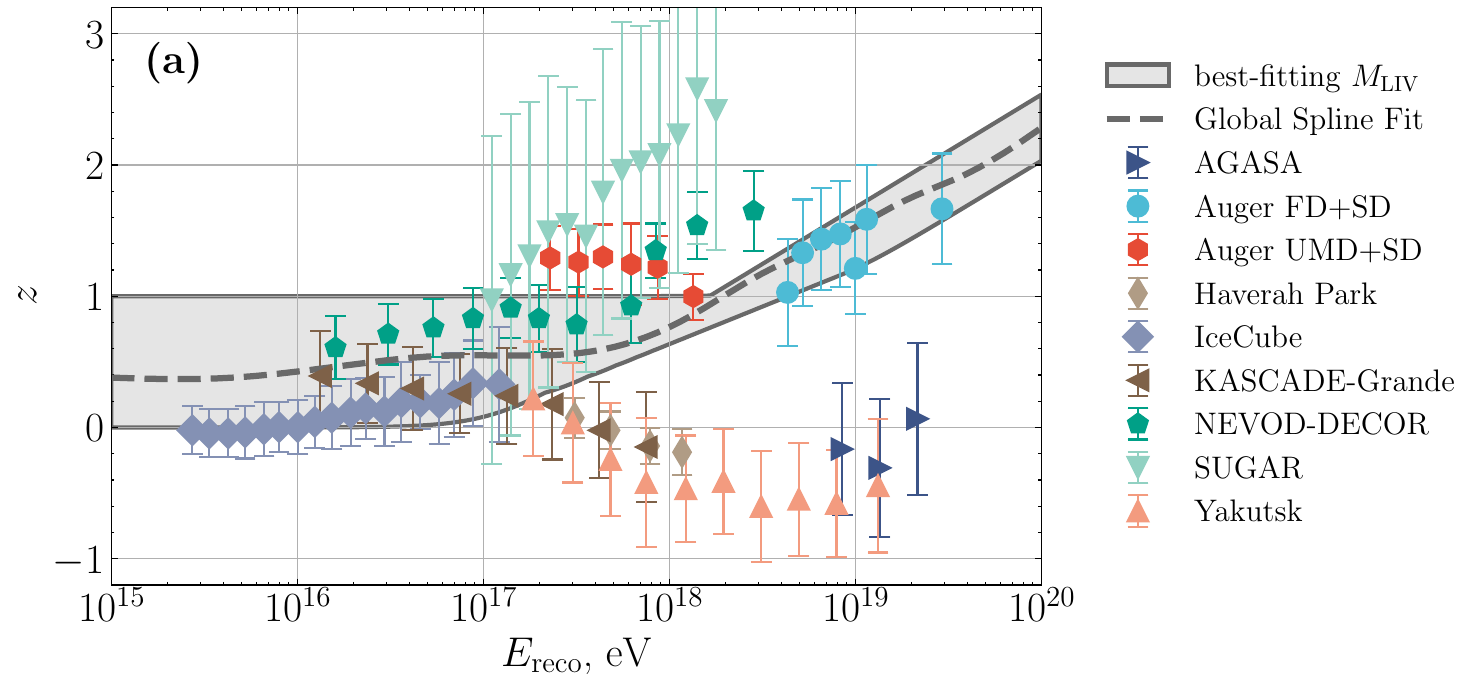}\\
    \vspace{\baselineskip}
    \includegraphics[width=0.95\columnwidth]{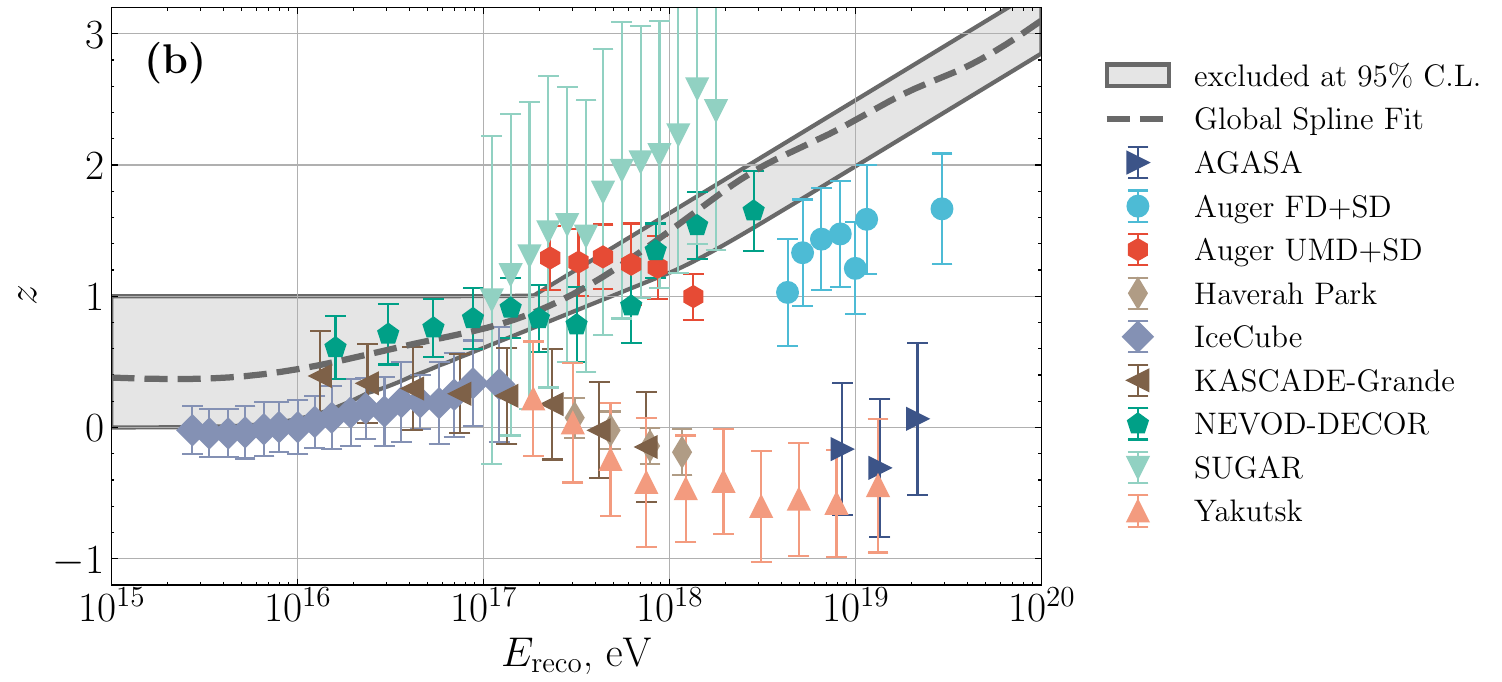}\\
    \vspace{\baselineskip}
    \includegraphics[width=0.95\columnwidth]{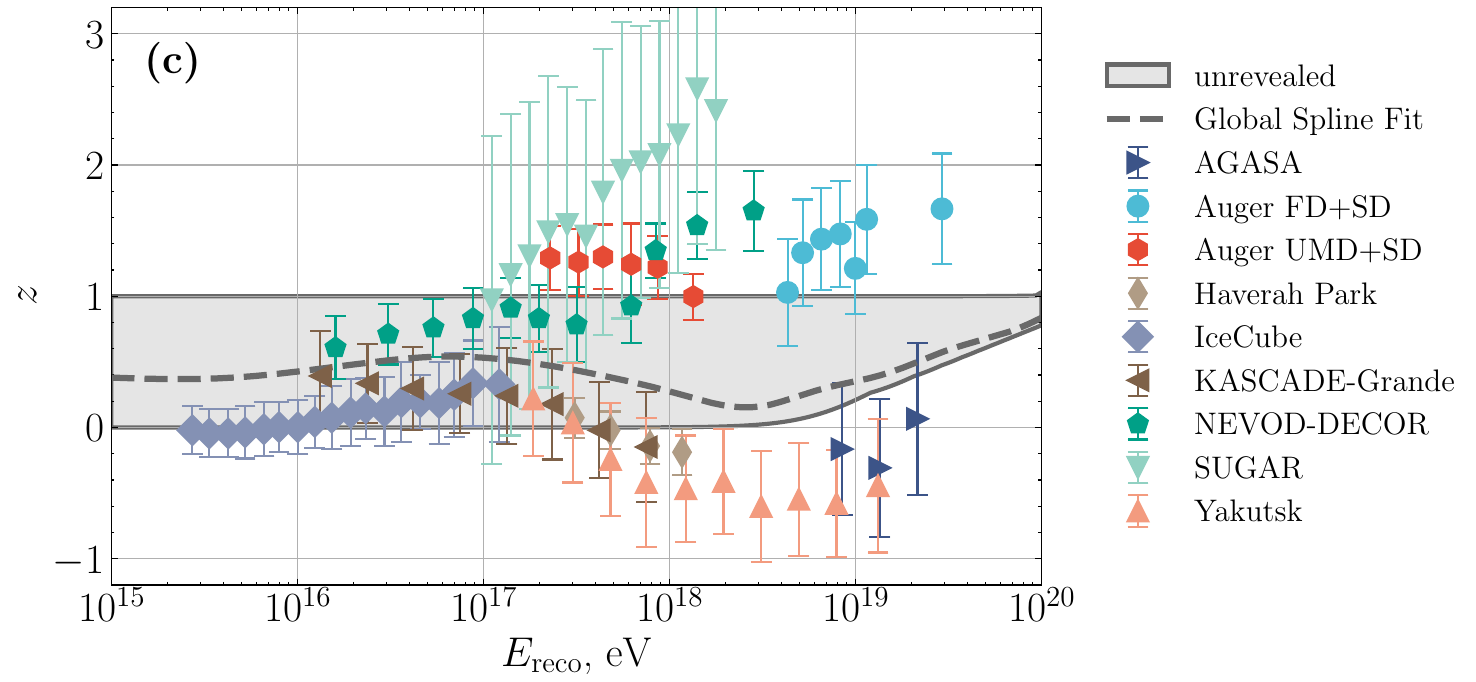}
    
    \caption{\(z\)-scale in experiments (see metaanalysis~\cite{ArteagaVelazquez:2023fda}) in comparison with the LIV scenario prediction. The shaded region corresponds to the LIV scenario \textbf{(a)} favored by the composition-independent maximum likelihood analysis, \(M_{\text{LIV}}=1.9\times 10^{16}\)~GeV; \textbf{(b)} excluded at \(95\%\) C.L. by the Auger data, \(M_{\text{LIV}}=2.4\times 10^{14}\)~GeV; \textbf{(c)} consistent with the LI scenario expectation, \(M_{\text{LIV}}=6.3\times 10^{19}\)~GeV. The dashed gray line assumes a composition from the Global Spline Fit~\cite{GSF_2017ICRC...35..533D}.}
    \label{fig:various-MLIV}
\end{figure}

The best-fitting LIV model prediction, in comparison with the relevant experimental data, is shown in Figure~\ref{fig:various-MLIV}, panel \textbf{(a)}. Additionally, panel~\textbf{(b)} and panel~\textbf{(c)}  illustrate, respectively, the predictions for \(M_{\text{LIV}}=2.4\times 10^{14}\)~GeV and for \(M_{\text{LIV}}=6.3\times10^{19}\)~GeV. The latter LIV scenario corresponds to \(p\)-value\,\(=0.33\).

\section{Conclusions}
\label{sec:concl}
We have shown that a subluminal LIV in the photon sector on the mass scale of \(M_{\text{LIV}} \sim 10^{16} \, \text{ GeV}\) could be an explanation for the muon puzzle. This parameter range has not yet been tested by data and observations. Moreover, \(M_{\text{LIV}} \sim 10^{(15\dots16)}\) GeV is favored in one of models of  Horava--Lifshitz gravity \cite{Blas:2010hb}. Additionally, subluminal quartic LIV in the photon sector can appear as a loop effect of LIV in other species \cite{Satunin:2017wmk}. 

The proposed LIV scenario can be falsified by measuring the ground-level muons spectrum. The latter must match that of a higher energy LI EAS.
The obtained numerical value of the LIV mass scale can be tested in a more direct way with the observation of air showers induced by primary photons of energy \(\sim 10^{(9\dots10)}\) GeV. Cosmogenic photons \cite{cosmogenic-photons} of these energies are produced by ultra-high-energy protons in the Greizen--Zatsepin--Kuzmin process \cite{G,ZK}. The corresponding analysis on LIV bounds has been performed in \cite{Rubtsov:2013wwa}. Thus, a detection of 
primary photons in this energy range would set a strong constraint $M_{\text{LIV}} > 10^{22}$ GeV, which may exclude the LIV explanation of the muon puzzle: this parameter range seems to have almost negligible effect to energy reconstruction.

Within the same approach, we set a primary-composition independent \(95\%\) C.L. bound on $M_{\text{LIV}}$ which reads $M_{\text{LIV}} > 2.4 \times 10^{14}$~GeV. 
The obtained constraint is an order of magnitude stronger than the current ones \cite{Satunin:2021vfx, Lang:2018yog}.

\section*{Acknowledgements}
We are indebted to N.\,N.~Kalmykov who emphasized the relevance of modeling the electromagnetic shower component to the muon puzzle. ST thanks his colleagues from the WHISP working group for illuminating discussions of the muon puzzle. 
NM and AS thank the Theoretical Physics and Mathematics Advancement Foundation “BASIS” for the student fellowships under the contracts 24-2-10-39-1 and 24-2-10-33-1, respectively. 

This work was supported by the Russian Science Foundation, grant 22-12-00253.
\vfill

\bibliography{citation}

\newpage
\onecolumngrid
\appendix
\vskip 8mm
\newpage
\centerline{\bf Supplementary Information}
\vskip 4mm
\twocolumngrid
\section{Electromagnetic cross sections in CORSIKA}
\label{sec:appendix:EM}
In CORSIKA 7~\cite{1998CORSIKA_Physics}, the electromagnetic interactions are modelled by means of EGS4 utilities~\cite{EGS4_Nelson1988}. Within this framework, the gamma-air interactions are divided into five types: photoelectric effect, Compton scattering, Bethe--Heitler process (that is, electron-positron pair production), photonuclear reaction, and muon pair production. Only the latter three types have non-vanishing contributions in the energy range of our interest.

The cross-sections of these reactions are encoded by their branching ratios,
\begin{equation}
    \operatorname{BR}_i(E_\gamma) = \frac{\sigma_i(E_\gamma)}{\sum\limits_{i'} \sigma_{i'}(E_\gamma)} = \frac{\sigma_i(E_\gamma)}{\sigma_{\text{tot}}(E_\gamma)},
\end{equation}
where the indices \(i, i'\) correspond to different processes, and \(E_\gamma\) is the photon energy in the laboratory frame. The total cross-section \(\sigma_{\text{tot}}(E_\gamma)\) is encoded by the photon mean free path in the atmosphere, \(\lambda_\gamma(E_\gamma) \propto \sigma_{\text{tot}}^{-1}(E_\gamma)\). 

During the runtime, the energy-dependent branching ratios and photon mean free path are calculated as follows:
\begin{equation}
    \label{eq:BR-lambda-in-CORSIKA}
    \begin{aligned}
    \lambda_\gamma &\propto \operatorname{GMFPR0},\\
    \operatorname{BR}_{\text{muon pair production}} &= \operatorname{GBR1},\\
    \operatorname{BR}_{\text{photonuclear reaction}} &= \operatorname{GBR2}-\operatorname{GBR1},\\
    \operatorname{BR}_{\text{photoelectric effect}} &= \operatorname{GBR3}-\operatorname{GBR2},\\
    \operatorname{BR}_{\text{Compton scattering}} &= \operatorname{GBR4}-\operatorname{GBR3},\\
    \operatorname{BR}_{\text{BH}} &= 1-\operatorname{GBR4}.\\
    \end{aligned}
\end{equation}
Within the Appendix, for clarity, we follow the notation used in the CORSIKA source code and CORSIKA User's Guide~\cite{corsika_user_guide}.
It is guaranteed that:
\begin{equation}
    \forall {\rm j}\leq {\rm j}',\quad 0 \leq \operatorname{GBRj} \leq \operatorname{GBRj}' \leq 1.
\end{equation}
\(\operatorname{GMFPR0}\) and \(\operatorname{GBRj}\) are computed as follows:
\begin{equation}
    \label{eq:arrays-in-CORSIKA}
    \begin{aligned}
    \operatorname{GMFPR0} &= \operatorname{GMFP0}_{\rm LGLE} + \operatorname{GMFP1}_{\rm LGLE} \times \operatorname{GLE},\\
    \operatorname{GBRj} &= \operatorname{GBRj0}_{\rm LGLE} + \operatorname{GBRj1}_{\rm LGLE} \times \operatorname{GLE},
    \end{aligned}
\end{equation}
where \(\rm j \in\{1, 2, 3, 4\}\), \(\rm LGLE\in\{1, \dots, NGE\}\) is the energy-dependent index, and \(\operatorname{GLE}\) is \(\ln[E_\gamma /1~\text{MeV}]\). The index \(\rm LGLE\) is calculated in a similar manner:
\begin{equation}
    {\rm LGLE} = \operatorname{GE0}+\operatorname{GE1} \times \operatorname{GLE},
\end{equation}
where rounding to the closest positive integer is assumed.

The arrays \(\operatorname{GMFP0}, \operatorname{GMFP1}, \operatorname{GBR10}, \dots, \operatorname{GBR41}\) of size \(\rm NGE\), as well as the variables \(\operatorname{GE0}\) and \(\operatorname{GE1}\) are all stored in data files \texttt{path/run/EGSDAT6\_x.x}, where \texttt{path} is the path to CORSIKA~7 installation directory, and \texttt{x.x} is the minimal 
kinetic energy to be followed in MeV.\vfill

\section{Implementation of the modified Bethe-Heitler cross section}
\label{sec:appendix:BH}
Let us introduce \(\xi_\gamma = E_\gamma (m_e M_{\text{LIV}})^{-1/2}\). To suppress the Bethe--Heitler process, we use a cross-section attenuation factor \(f(\xi_\gamma)=\sigma^{\text{LIV}}_{\text{BH}}/\sigma^{\text{LI}}_{\text{BH}}\). Eq.~\eqref{eq:LIV-suppression} determines \(f(\xi_\gamma)\) in the region of \(\xi_\gamma \gg 1\). In the opposite case, when \(\xi \ll 1\), \(f(\xi_\gamma) = 1\). In the intermediate region, we manually adjust \(\tanh\)-like smooth transition between the two asymptotic regimes.

The modified total cross-section:
\begin{equation}
    \begin{aligned}
    \sigma^{\text{LIV}}_{\text{tot}} &= \sigma_{\text{BH}}^{\text{LIV}} + \sum\limits_{i'\neq \text{BH}} \sigma_{i'}= (f-1)\sigma_{\text{BH}}^{\text{LI}} + \sum\limits_{i'} \sigma_{i'} =\\
    &=\sigma_{\text{tot}}^{\text{LI}} \times \left[(f-1)\operatorname{BR}_{\text{BH}}^{\text{LI}}+1\right]
    \end{aligned}
\end{equation}
Since \(\lambda_\gamma \propto \sigma_{\text{tot}}^{-1}\), the modified photon mean free path:
\begin{equation} 
    \lambda_\gamma^{\text{LIV}} / \lambda_\gamma^{\text{LI}} = \frac{1}{ (f-1)\operatorname{BR}_{\text{BH}}^{\text{LI}}+1}
\end{equation}
Finally, the modified branching ratios:
\begin{equation}
    \begin{aligned}
        \operatorname{BR}_{\text{BH}}^{\text{LIV}}/\operatorname{BR}_{\text{BH}}^{\text{LI}} &= \frac{f}{(f-1)\operatorname{BR}_{\text{BH}}^{\text{LI}} + 1},\\
        \operatorname{BR}_{i'}^{\text{LIV}}/\operatorname{BR}_{i'}^{\text{LI}} &= \frac{1}{(f-1)\operatorname{BR}_{\text{BH}}^{\text{LI}} + 1},~i'\neq \text{BH}.
    \end{aligned} 
\end{equation}
Using Equations~\eqref{eq:BR-lambda-in-CORSIKA}  and \eqref{eq:arrays-in-CORSIKA}, we induce the corresponding transformations of the EGS4 arrays \(\operatorname{GMFP0}, \operatorname{GMFP1}, \operatorname{GBR10}, \dots, \operatorname{GBR41}\). We emphasize that our code directly overwrites \texttt{EGSDAT6\_x.x} files without changing anything in the CORSIKA source code. Once these files are overwritten, the desired LIV model is implemented to EGS4.
\end{document}